\documentclass[a4paper,amssymb,amsmath,pra,twocolumn,floatfix,showpacs]{revtex4}
\usepackage{enumerate,
units,
fancyhdr,
amsmath,
amssymb,
graphicx,
color}

\DeclareGraphicsExtensions{.png,.eps}

\newcommand{\ket}[1]{|#1 \rangle}
\newcommand{\bra}[1]{\langle #1|}
\newcommand{\braket}[2]{\langle #1| #2\rangle}

\begin{document}

\title{Digital three-state adiabatic passage}

\author{\firstname{Jesse} A. \surname{Vaitkus}}
\affiliation{Applied Physics, School of Applied Sciences, RMIT University, Melbourne 3001, Australia}
\author{\firstname{Andrew} D. \surname{Greentree}}
\affiliation{Applied Physics, School of Applied Sciences, RMIT University, Melbourne 3001, Australia}
\email{andrew.greentree@rmit.edu.au}

\begin{abstract}
We explore protocols for three-state adiabatic passage where the tunnel matrix elements are varied digitally, rather than smoothly as is the case with conventional adiabatic passage.  In particular, we focus on the STIRAP and related three-state schemes where the control is applied stepwise, with either equal spaced levels for the tunnel matrix elements or uniform pulse lengths.  Our results show that the evolution typically shows the hallmarks of conventional adiabatic passage, although with additional resonances exhibiting no state transfer.
\end{abstract}

\pacs{05.60.Gg, 42.50.Ex, 03.67.-a, 03.65.Aa}

\keywords{quantum}

\maketitle

\section{Introduction}

Adiabatic passage is an interesting and important tool for manipulating quantum states.  With adiabatic processes, one attempts to find a control strategy that maintains the system in an instantaneous eigenstate, whilst that Hamiltonian is smoothly varied from some initial state to the desired final state.  Such control strategies have been used and proposed for a range of tasks including the manipulation of population within an atomic system ~\cite{bib:GRB+1988,bib:KTS2007}, control of chemical reactions~\cite{bib:KR1998}, control of vibrational population~\cite{bib:CK1995}, and adiabatic quantum computation~\cite{bib:FGG+2001}.  

\begin{figure}[bh!]
\includegraphics[width=0.9\columnwidth,clip]{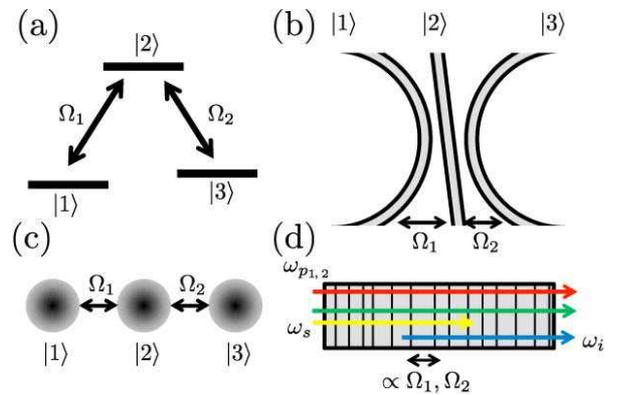}
\caption{Three-state adiabatic passage has been explored in many systems with similar Hamiltonians, and four examples are shown here.  (a) STIRAP effects state transfer between states of a three-level atom, typically ground states of a $\Lambda$ system, where the tunnel matrix elements are the Rabi frequencies of control lasers. (b) A three-waveguide system can be used to effect CTAP variation where the tunnel matrix elements are varied due to the evanescent coupling between waveguides (c) CTAP in a triple-dot or triple-donor system is effected by gate-controlled variations in the wave function overlap of the particle between the sites. (d) For nonlinear frequency conversion, the required coupling variation is achieved by superimposing two chirped poling functions in a periodically poled nonlinear material, assisted by two strong pump fields.}
\label{fig:analog}
\end{figure}

One of the tasks that adiabatic passage is particularly suited for is state transfer.  State transfer is an important task in its own right, but also serves as a useful metric for evaluating new techniques.  Here we focus explicitly on the task of state transfer in three-state systems using the STIRAP (STImulated Raman Adiabatic Passage) \cite{bib:KTS2007} and related protocols such as Coherent Tunneling Adiabatic Passage (CTAP) \cite{bib:ELR+2004,bib:GCH+2004,bib:SB2004,bib:P2006,bib:GKW2006,bib:LDO+2007,bib:RCP+2008}, Dark State Adiabatic Passage (DSAP) \cite{bib:OEO+2008,bib:OSF+2013}, and adiabatic nonlinear frequency conversion \cite{bib:PA2012}, see Fig.~\ref{fig:analog}.  In these schemes we have an effective three-state system where the states are denoted $\ket{1}$, $\ket{2}$, $\ket{3}$ with nearest neighbour tunnel matrix elements $\Omega_1$ between states $\ket{1}$ and $\ket{2}$, and $\Omega_2$ between states $\ket{2}$ and $\ket{3}$.  The task of these protocols is to effect population transfer from state $\ket{1}$ to $\ket{3}$ via the dark state, or its equivalent, $\ket{\mathcal{D}_0} = \left(\Omega_{2}\ket{1} - \Omega_1\ket{3}\right)/\sqrt{\Omega_1^2 + \Omega_2^2}$.  Our aim here is to discover the limitations of employing piecewise constant, i.e. digital, tunnel matrix elements, rather than the smoothly varying control that is typically thought to be required.  

Piecewise adiabatic passage (PAP) was introduced by Shapiro \textit{et al.} \cite{bib:SMM+2007} in the context of femtosecond pulse control of the internal states of atomic systems.  In this case the control fields are applied for short periods of time with an overall envelope that ensures the evolution.  Although formally this scheme cannot be considered as truly adiabatic due to the pulsed control, when viewed in the frequency domain, and observing that population cannot leak between eigenstates when the control fields are off, \cite{bib:SMS2009}, adiabatic-like evolution can be obtained.  Controls comprising sequences of short pulses have been explored quite extensively recently in the context of PAP, see for example \cite{bib:TGV2011,bib:RV2012}. The limits to such evolution in a general setting have also been explored by Boixo and Somma \cite{bib:BS2010}.  

Here we show the effects of employing a digital variation in the control fields, i.e. where the tunnel matrix elements are forced to vary in discrete steps, rather than smoothly.  This case is especially important for adiabatic frequency conversion using periodically poled media \cite{bib:PA2012}, and the inevitable digitisation that arises from the use of digital to analog converters for electronic control signals.  It is also likely that many computational models that use some form of finite element method in solving adiabatic evolution show some \textit{de facto} digitisation of the style we consider here.  Our results show that digital control can yield state transfer approximating more conventional three-state adiabatic passage, although we identify resonances where the transport does not exhibit adiabatic-like evolution. 

This paper is organised as follows.  We first introduce the Hamiltonian and control scheme for the digital adiabatic passage (DAP).  We then provide a simple analytic overestimate of the fidelity of DAP.  We consider two means for digitising the adiabatic passage, with uniform spacing of the pulses in time and uniform distribution of the tunnel matrix elements.  Both of these cases are investigated, and a method for generating DAP schemes where there is uniform spacing in the control parameter is also discussed.  Our results show adiabatic-like evolution with resonances of no evolution, and these resonances are used to identify optimal evolution strategies within the DAP framework. 

\section{Hamiltonian}

We consider a three-state system described by a tight-binding model, where each state is coupled only to its neighbour.  This model is useful for exploring STIRAP-like processes, although extensions to more realistic Hamiltonians have also been considered, especially in the context of spatial variations (i.e. CTAP), see for example \cite{bib:CGH+2008,bib:RPC+2009,bib:RCP+2008}.  Although these adiabatic processes only require energy degeneracy between the end states, for simplicity we assume that all states have the same energy.  In this case the Hamiltonian becomes (with $\hbar = 1$, and $\Omega_1$ and $\Omega_2$ real)
\begin{align}
\mathcal{H} = \Omega_1 \left(\ket{1}\bra{2} + \ket{2}\bra{1}\right) + \Omega_2 \left(\ket{2}\bra{3} + \ket{3}\bra{2}\right).
\end{align}
Note that $\Omega_1$ and $\Omega_2$ are assumed to vary according to some time-varying external control, as in the case of STIRAP and CTAP, or spatially varying design, as in the case of adiabatic passage in waveguides or nonlinear frequency conversion with periodically poled materials.  For simplicity and without loss of generality, we will assume that the $\Omega$ vary as a function of time.

To understand the adiabatic evolution of the three-state system, we first solve for the eigenvectors of the Hamiltonian, which are
\begin{subequations}
\begin{align}
|\mathcal{D}_\pm\rangle = &\frac{\Omega_{1}|1\rangle \pm \sqrt{\Omega_{1}^2+\Omega_{2}^2}|2\rangle + \Omega_{2} |3\rangle}{\sqrt{2(\Omega_{1}^2 + \Omega_{2}^2)}} ,\\
|\mathcal{D}_0\rangle = &\frac{\Omega_{2}|1\rangle - \Omega_{1}|3\rangle}{\sqrt{\Omega_{1}^2+\Omega_{2}^2}},
\end{align}
\end{subequations}
with corresponding eigenenergies
\begin{align}
E_{\pm} = \pm \sqrt{\Omega_{1}^2+\Omega_{2}^2} = \pm \mathcal{E}, \quad E_0 =  0.
\end{align}
Here $\ket{\mathcal{D}_0}$ is usually referred to as the dark state in the context of STIRAP, or the null state for CTAP.  

\begin{figure}[tb!]
\includegraphics[width = 0.95\columnwidth, clip]{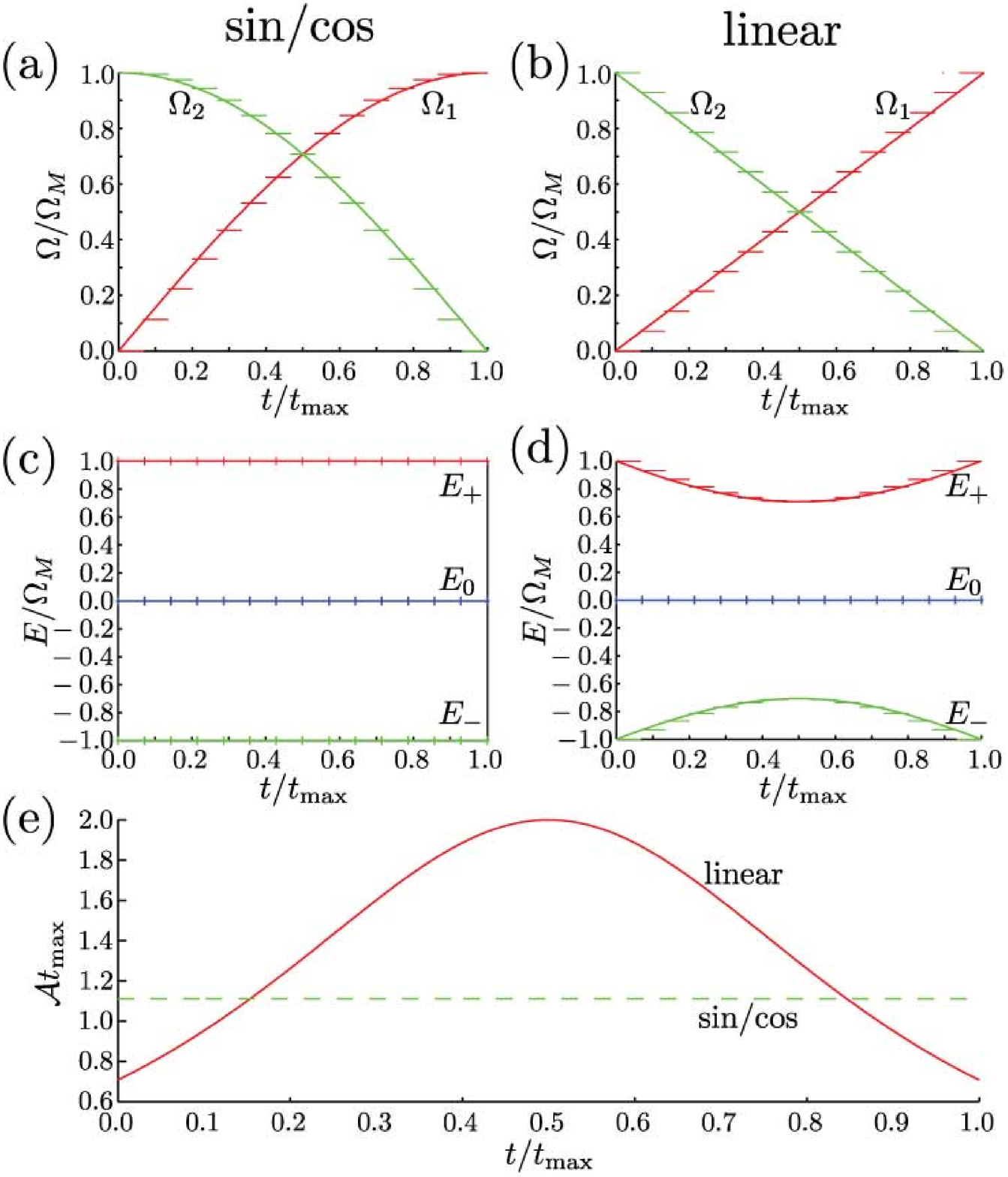}
\caption{Coupling schemes and eigenspectra for the case of sin/cos and linear coupling. (a) sin/cos coupling scheme showing both the continuous and digital variation with 15 steps. (b) linear coupling scheme with showing both continuous  and digital coupling for 15 steps. (c) Eigenspectrum for the sin/cos scheme.  Note that the eigenvalues do not change throughout the protocol although the state compositions do change for each set of tunnel matrix elements. (d) Eigenspectrum for the linear coupling scheme, showing that the point of minimum energy separation is at the midpoint of the protocol.  This is also the point where the adiabaticity parameter is maximised in the continuous version. (e) Adiabaticity for the sin/cos (green dashed line) and linear (red solid line) protocols as a function of fractional time.  Here we normalise the adiabaticity by multiplying by the total time.  The sin/cos protocol is a straight line, whereas the linear protocol shows the more standard maximum in the adiabaticity at the midpoint of the evolution.}
\label{fig:couple}
\end{figure}

For adiabatic passage, the strategy to realise state transfer from $\ket{1}$ to $\ket{3}$ is relatively straightforward.  One typically chooses some smoothly-varying control function for $\Omega_1$ and $\Omega_2$, with the constraints that $\Omega_2(t = 0) \gg \Omega_1 (t=0)$ and $\Omega_1(t = t_{\max}) \gg \Omega_2 (t=t_{\max})$, and the time $t$ varies from 0 to $t_{\max}$, where there is considerable overlap of the pulses for all intervening times.  The functional form is relatively unimportant and many different pulse sequences have been employed.  Here we employ a sinusoidal control sequence \cite{bib:CH1990,bib:LS1996} such that
\begin{align}
\Omega_{1} = \Omega_M \sin\left(\frac{t \pi}{2 t_{\max}}\right), \quad 
\Omega_{2} = \Omega_M \cos\left(\frac{t \pi}{2 t_{\max}}\right), \label{eq:sincos}
\end{align}
as shown in Fig.~\ref{fig:couple}(a), where $\Omega_M$ is the maximum tunnel matrix element.  This control sequence has the advantage that none of the eigenvalues of the system vary during the adiabatic passage, i.e. $E_{\pm}(t) = \pm\Omega_M$ [Fig.~\ref{fig:couple}(c)].  

The adiabaticity parameter is a measure for how well a system stays in an instantaneous eigenstate as the Hamiltonian is varied \cite{bib:Mes}.  So for any two instantaneous eigenstates $\psi_1$ and $\psi_2$ we define
\begin{align}
\mathcal{A} \equiv \frac{\bra{\psi_1} \partial_{t} \mathcal{H} \ket{\psi_2}}{|\bra{\psi_1}\mathcal{H}\ket{\psi_1} - \bra{\psi_2}\mathcal{H}\ket{\psi_2}|^2},
\end{align}
and for adiabatic evolution, we require $\mathcal{A} \ll 1$.  For the pulse sequences typically employed for STIRAP/CTAP, the adiabaticity peaks either at or near the midpoint of the transfer \cite{bib:DP1976} (for a comparison of different schemes see \cite{bib:CGH+2008}). For the case in eq.~\ref{eq:sincos}, however, the adiabaticity parameter is constant, namely
\begin{align}
\mathcal{A}_{\mathrm{sc}} = \frac{\pi \sqrt{2}}{4 t_{\max} \Omega_M}.
\end{align}

We also consider a linear variation in the couplings [Fig.~\ref{fig:couple}(b)], 
\begin{align}
\Omega_1 = \Omega_M (t/t_{\max}), \quad \Omega_2 = \Omega_M (1 - t/t_{\max}).
\end{align}
This scheme has the advantage of simplicity and also corresponds to systems where the control levels are equally spaced, but this scheme is less desirable for high fidelity transport, as will be shown below.  The adiabaticity in the linear case is
\begin{align}
\mathcal{A}_{\mathrm{lin}} = \frac{1}{4 t_{\max} \Omega_M} \left[\left(\frac{t}{t_{\max}}\right)^2- \frac{t}{t_{\max}}+\frac{1}{2}\right]^{-3/2}.
\end{align}
In Fig.~\ref{fig:couple} we compare the control sequences, eigenenergies and adiabaticity (in the smoothly varying case) for the two coupling schemes.  Note that the linear evolution shows the point of minimum energy separation at the midpoint of the protocol, in keeping with most of the other pulsing schemes that have been considered in the literature to date, whilst the eigenergies are constant for the sin/cos sequence.  The fact that the adiabaticity for the linear sequence is larger for the same total time shows immediately that the linear scheme should be less efficient for state transfer than the sin/cos scheme.  The constancy of the eigenenergies in the sin/cos scheme also has benefits for the DAP, as discussed below.


\section{Equal pulse length digitisation}

The aim of this work is to understand the consequences of digital control on three-state adiabatic passage protocols.  It is important to realise that the concept of adiabaticity is formally inapplicable in this case, as the assumption of ideal digital control implies step function variation in the control parameters.  Hence adiabatic following is not strictly possible.  Nevertheless, as we will show, the evolution obtained by a na\"{\i}ve digitisation of the controlling tunnel matrix elements provides evolution that typically mimics the adiabatic evolution found for smoothly varying control fields, as was the case with PAP~\cite{bib:SMM+2007}.

For digital control, the tunnel matrix elements are piecewise constant with time.  For $N$ steps, we use
\begin{align}
\Omega_{1} = \Omega_M \sin\left[\frac{\xi \pi}{2(N-1)}\right], \quad
\Omega_{2} = \Omega_M \cos\left[\frac{\xi \pi}{2(N-1)}\right],
\end{align}
where
\begin{align}
\xi = \left\lfloor \frac{N t}{t_{\max}} \right\rfloor,
\end{align}
and $t \in  [0,t_{\max})$.

If the evolution were adiabatic, then we could use adiabaticity arguments to determine the overlap of any given state with the eigenstate of the infinitesimally incremented Hamiltonian.  However, the digitisation introduces a series of discrete jumps.  To determine an estimate of the error, we therefore look at the overlap between the null states before and after a change in the Hamiltonian, reasoning that in the limit of infinitely many steps that the continuously varying Hamiltonian should be approximated. Due to the simple form of the digital control, the difference between the $\xi^{\mathrm{th}}$ and $(\xi+1)^{\mathrm{th}}$ null state is
\begin{align}
\eta(\xi) &= 1 - \left|\braket{\mathcal{D}_0(\xi)}{\mathcal{D}_0(\xi+1)}\right|^2 \nonumber \\ &= \sin^2 \left[ \frac{\pi}{2 (N-1)} \right]. \label{eq:ovlap}
\end{align}
In the limit that $\eta \ll 1$, the total error in the DAP is the sum over all the individual errors
\begin{align}
\eta_T = \sum\limits_{\xi=0}^{N-1} \eta(\xi) = N \sin^2 \left[ \frac{\pi}{2 (N-1)} \right], \label{eq:AdiabaticError}
\end{align}
which becomes in the limit of large $N$
\begin{align}
\eta_T = \pi^2/(4N). 
\end{align}

The above approximation gives a quick measure of the fidelity of the digital adiabatic process, confirming the intuition that  increasing the number of steps should improve the overall fidelity.  However, it misses some extremely important physics, in particular the fact that population may be projected back into the null state from the other two eigenstates~\cite{bib:VCM+2012}, and also the fact that there will be interference in the accumulated phases of the residual population.  Hence the error determined in eq.~\ref{eq:AdiabaticError} is in a real sense `timeless'.  The $\sin$/$\cos$ evolution is particularly useful for understanding the evolution, as all eigenstates have constant energy throughout the protocol, although of course they change their exact composition.

\begin{figure*}[tb!]
\includegraphics[width=0.75\textwidth,clip]{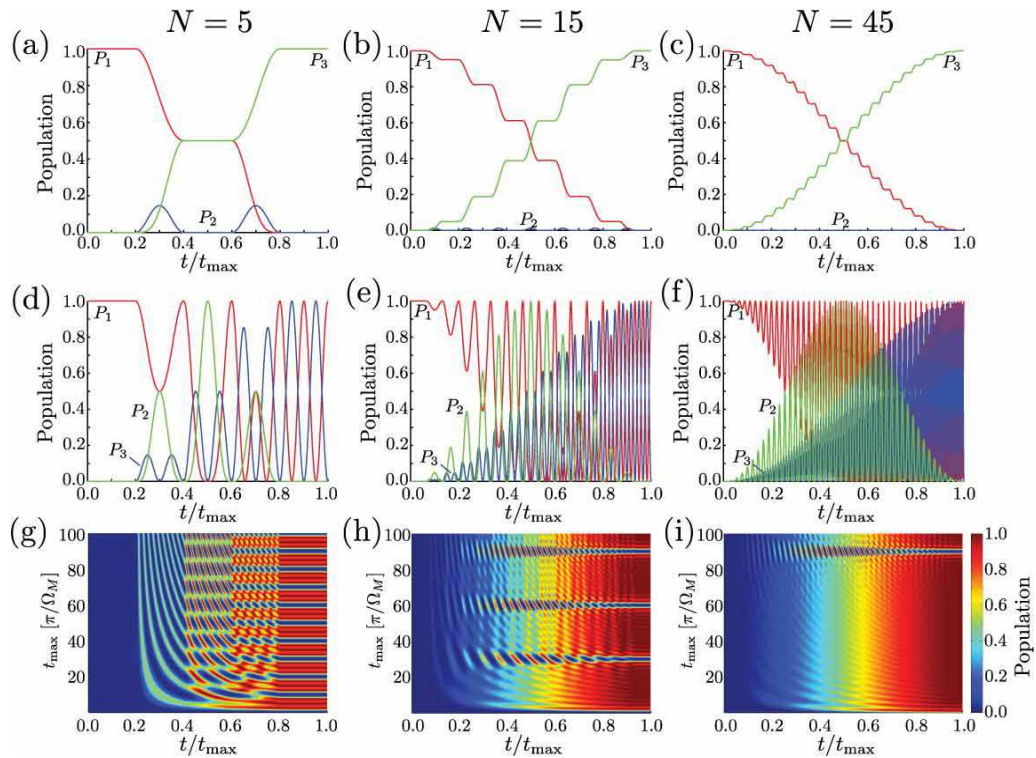}
\caption{Temporal evolution through DAP.  Adiabatic-like evolution as a function of fractional time for (a) $N = 5$, $t_{\max} = 5\pi/\Omega_M$; (b) $N = 15$, $t_{\max} = 15\pi/\Omega_M$; and (c) $N = 45$, $t_{\max} = 45\pi/\Omega_M$.  The evolution is highly reminiscent of conventional three-state adiabatic passage, with the similarities becoming greater with increasing number of steps. When the time per step is $\tau = 2n\pi/\Omega_M$ then the evolution operator becomes the identity at the end of each step, hence the \emph{total} evolution is the identity operator.  This is illustrated for the cases that (d) $N = 5$, $t_{\max} = 10\pi/\Omega_M$; (e) $N = 15$, $t_{\max} = 30\pi/\Omega_M$; and (f) $N = 45$, $t_{\max} = 90\pi/\Omega_M$. Pseudocolor plots showing the evolution as a function of fractional time and total time for (g) $N = 5$, (h) $N=15$ and (i) $N=45$.  Notice that the evolution is mostly adiabatic-like, with the addition of the resonances where the evolution is the identity.}
\label{fig:evolution}
\end{figure*}

\begin{widetext}
To perform accurate numerical calculations, we generate the unitary evolution operator corresponding to DAP, which is straightforward due to the piecewise constant nature of the digital Hamiltonian. The unitary evolution operator is constructed so that $\ket{\psi(\xi t_{\max}/N)} = U(\xi) \ket{\psi[(\xi-1)t_{\max}/N]}$.  Note that for the $\sin$/$\cos$ sequence the energy of all eigenstates is constant and so in this case $\mathcal{E} = \Omega_M$.
\begin{align}
U (\xi)  = \frac{1}{\mathcal{E}^2}\left[ \begin{array}{ccc}
\Omega_2^2 + \Omega_1^2 \cos (\mathcal{E} \tau) & i \mathcal{E} \Omega_1 \sin (\mathcal{E}  \tau) & \Omega_1\Omega_2 \left[ \cos(\mathcal{E} \tau) -1 \right] \\
i \mathcal{E} \Omega_1 \sin (\mathcal{E} \tau) & \mathcal{E}^2 \cos (\mathcal{E} \tau) & i \mathcal{E} \Omega_2 \sin (\mathcal{E} \tau) \\
\Omega_1\Omega_2 \left[\cos(\mathcal{E} \tau)-1\right] & i \mathcal{E} \Omega_2 \sin (\mathcal{E} \tau) & \Omega_1^2 + \Omega_2^2 \cos (\mathcal{E} \tau)
 \end{array}\right],
\end{align}
where we have introduced $\tau$, the length of the evolution under the current Hamiltonian. For equally spaced time steps $\tau = t_{\max}/N$.  The total evolution over the DAP is
\begin{align}
\mathcal{U} = U(N) U(N-1) \cdots U(\xi) \cdots U(2) U(1).
\end{align}
\end{widetext}

We first illustrate the process by showing several instances of the evolution for varying number of steps and total time.  Fig.~\ref{fig:evolution} shows the evolution of the populations, $P_i = |\braket{i}{i}|^2$, for $N = 5$, $10$, and $15$.  First considering the evolution through the protocol for varying $t_{\max}$ [Figs.~\ref{fig:evolution} (g) (h) and (i)] we notice evolution that is strongly reminiscent of adiabatic passage, with fairly smooth increase in population in $\ket{3}$.  However there are also pronounced resonances where the final population in $\ket{3}$ is zero.  The separation (in terms of $t_{\max}$) between these resonances increases as the number of digital steps is increased, and this result will be explained below.  

Figures~\ref{fig:evolution} (a) (b) and (c) show examples of adiabatic-like evolution that is indicative of most of the parameter space.  These results show that although the DAP does not show true adiabaticity, nevertheless the operational advantages of adiabatic passage (robustness against errors, smooth population variation and minimisation of population in the intervening state) are all preserved, with increasing number of steps improving the approximation to true adiabatic processes.  Notice that the maximum population in $\ket{2}$ is given by $\eta(\xi)$ from Eq.~\ref{eq:ovlap}.

The resonances of low-fidelity transport are explored in Figs.~\ref{fig:evolution} (d) (e) and (f), which show evolution through the centre of the resonances.  Here we observe complex behaviour that does not transfer population to $\ket{3}$.  We understand these resonances by noting that when $t_{\max} = 2nN\pi/\Omega_M$, with $n \in \mathbb{N}^0$ in each case, then $\tau = 2n\pi/\Omega_M$ and $U(\xi)$ is the identity matrix.  Hence the overall evolution must also be the identity.

\begin{figure}[tb!]
\includegraphics[width = 0.95\columnwidth, clip]{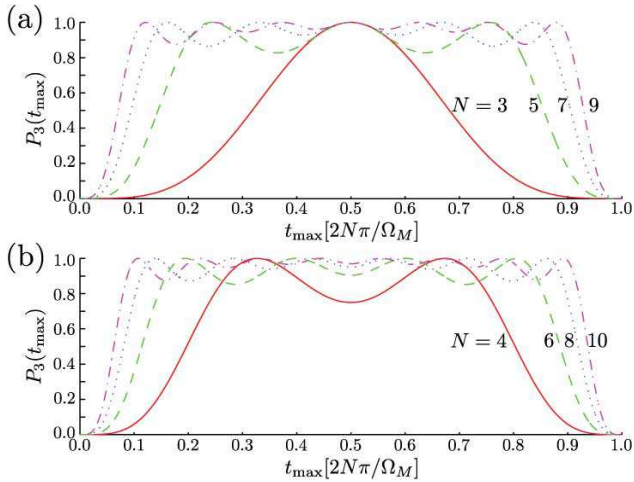}
\caption{Final population in $\ket{3}$ after DAP using the sin/cos protocol as a function of $t_{\max}$ for the first four non-trivial $N$ for (a) $N$ odd ($N = 3$, $5$, $7$, and $9$) and (b) $N$ even ($N = 4$, $6$, $8$, and $10$).  When $\cos(\Omega_M \tau) = 1$, then the evolution becomes the identity, which is seen by the repeating nulls in the population.  Also of interest are the local maxima in population transfer, which correspond to the periodic Dirichlet kernel.   As the number of digital steps increases, the number of local maxima increase and the pattern approaches a square function.  We have not shown the corresponding figures for the linear case as the results are less insightful.}
\label{fig:dirichlet}
\end{figure}

In Fig.~\ref{fig:dirichlet} we plot the population in state $\ket{3}$ at the end of the DAP protocol as a function of $t_{\max}$ for different values of $N$.  The resonances corresponding to when $\mathcal{U} = I$ are seen by the dips as a function of $t_{\max}$.  Also of interest are the local maxima and minima in population transfer between these resonances.  We find that the pattern of these extrema phenomenologically corresponds to a subset of the periodic Dirichlet kernel, however the heights of the nodes did not.  We also note that protocols with odd $N$ and total time of $N \pi$, exhibit comparable evolution to those reported in Ref.~\cite{bib:RV2012}.  The elegance of the results for the $\sin/\cos$ scheme are not replicated with other protocols for the timing scheme utilised.

\begin{figure}[tb!]
\includegraphics[width=0.7\columnwidth,clip]{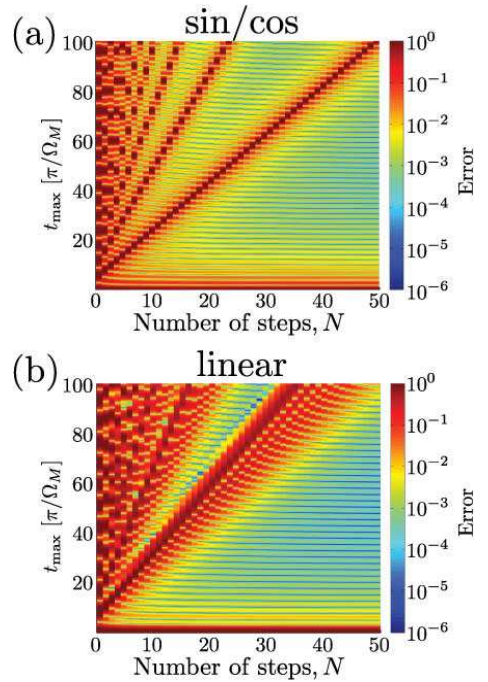}
\caption{Transport error $[1 - |\braket{\psi(t_{\max})}{3}|^2]$ as a function of $N$ and $t_{\max}$, for (a) the sin/cos scheme, and (b) the linear scheme.  There is relatively low error away from the identity evolution resonances and with increasing N the spacing of these resonances increases linearly as does the number of optimal nodes. In each case, the horizontal axis begins at N=3 as this is the first non-trivial result.  Note that the sin/cos scheme has sharper resonances as the eigenergies are constant throughout the protocol, whereas the changing energy observed in the linear scheme leads to a band over which some of the evolutions are the identity.}
\label{fig:p3-error}
\end{figure}

We further explore the fidelity by showing in Fig.~\ref{fig:p3-error} the error in DAP transfer as a function of $N$ and $t_{\max}$ for both the sin/cos and linear schemes.  The traces in Fig.~\ref{fig:dirichlet} correspond to vertical slices through Fig.~\ref{fig:p3-error} (a).  Again, noticeable are the resonances showing identity evolution, and also apparent are the local maxima identified above.  For the linear scheme [Fig~\ref{fig:p3-error} (c)], the resonances are more complicated and do not result in a perfect identity.  This is because $\Omega_1^2 + \Omega_2^2$ is no longer constant, and hence perfect identity evolution across every digital step is not possible for a given $\tau$.  


\begin{figure}[tb!]
\includegraphics[width=0.95\columnwidth,clip]{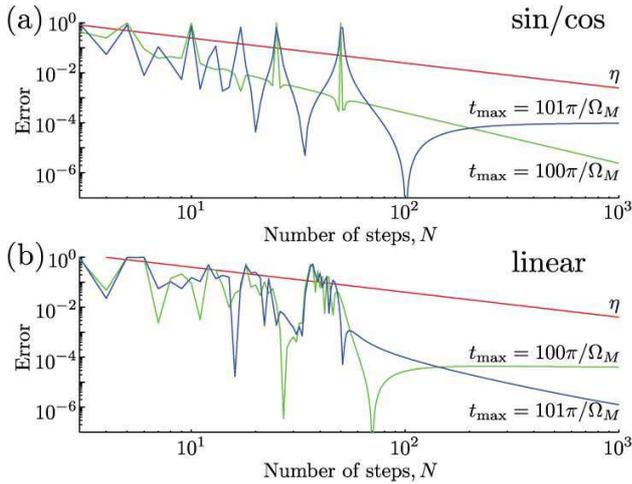}
\caption{Transport fidelity at  $t_{\max} = 100\pi/\Omega_M$ (green) and $t_{\max} = 101\pi/\Omega_M$ (blue) for varying number of steps, $N$, compared with the analytical results for $\eta$ (red). (a) Shows results for the sin/cos scheme, and (b) shows the case for the linear scheme.  As expected, as $N$ increases, the fidelity improves, except for resonance points as discussed above.  Notice the difference between the even and odd results, which is more pronounced for the sin/cos scheme than the linear scheme.}
\label{fig:error}
\end{figure}

We also compare the final state population for even and odd $N$ with the analytical estimate from Eq.~\ref{eq:AdiabaticError} for increasing $N$ in Fig.~\ref{fig:error}.  The analytic estimate does not capture the resonant phenomena, but does show the increasing fidelity of DAP with increasing $N$.  In both the linear and sin/cos schemes we see a general trend in the reduction of error with increasing $N$ and a difference between evolution where the total time is an even or odd multiple of $\pi/\Omega_M$.  However, as was evident in Fig.~\ref{fig:p3-error}, the poor fidelity resonances are less clear in the linear scheme than for the sin/cos scheme.  This result is a consequence of the fact that the eigenspectrum has different values as a function of fractional time in the linear scheme, and hence the resonances are broader and less pronounced.

\section{Uniformly varying tunnel matrix elements, non-uniform pulse length}


In the previous section we considered the simplest approach with respect to timing; the pulses defining the protocol were designed to have equal temporal extent.  Although conceptually straightforward, in  some situations the control over the timing of pulses may be greater than the control over the level of the pulses, and hence for such situations it is more natural to consider digital schemes where the pulse length is optimised for a given digitisation level.  As the ultimate limit of such control imperatives, we may consider bang-bang control as a one-bit control scheme with fine timing control \cite{bib:VL1998}.  For simplicity we consider only the case where the tunnel matrix element is varied between uniformly spaced levels, however our method may be easily generalised to the case where the parameter controlling the tunnel matrix element (e.g. a voltage level for gate-driven processes or waveguide separation for waveguide adiabatic passage) is the uniformly spaced digital quantity.

\begin{figure}
\includegraphics[width=0.95\columnwidth,clip]{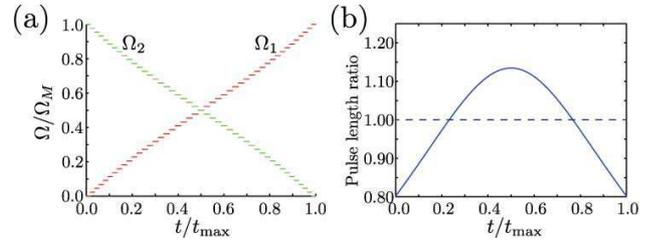}
\caption{(a) Linear scheme with energy-dependent pulse length.  The tunnel matrix elements are taken from a set of equally spaced levels and the pulse time is inversely proportional to the energy. (b) Variation in pulse length relative to the uniformly spaced scheme. More time is spent in lower energy areas than high meaning that the pulse length at the midpoint of the protocol is longer than at any other time in the protocol.}
\label{fig:LinearOptimisedTime}
\end{figure}
 
To design the appropriate pulse length, recall that for a given digital step, identity evolution occurs when the pulse length is $\tau = 2\pi/\mathcal{E}$.  Optimal complete transfer occurs for step $\xi$ when the length corresponds to 
\begin{align}
\tau_{\xi} = \frac{\pi}{\mathcal{E}}, \label{eq:OptimalTime}
\end{align}
where $\mathcal{E} = \mathcal{E}(\xi)$.  More generally, one can determine an equivalent to Fig.~\ref{fig:dirichlet} for the compensated linear scheme, which is stretched by a factor dependent on the number of steps and the precise details of the scheme used.

The compensation scheme described above provides a method to determine the pulse length for \emph{any} control protocol where the $\mathcal{E}$ are known.  To illustrate this method for setting the pulse length, we return to the linear control scheme where we require the tunnel matrix elements to be varied between equally spaced levels.  However we now optimise the time according to eq.~\ref{eq:OptimalTime}.  This modified control scheme is illustrated in Fig.~\ref{fig:LinearOptimisedTime}, which shows the control protocol as well as the shift in the pulse length relative to the uniform pulse time protocol discussed above.  The optimal pulse time approach means that the pulse length at the midpoint of the protocol is longer than at the beginning or end of the protocol.  Interestingly, this is a similar result to that obtained from standard adiabaticity analysis, although the reason for longer pulses is different in this case.





\begin{figure}
\includegraphics[width=0.95\columnwidth,clip]{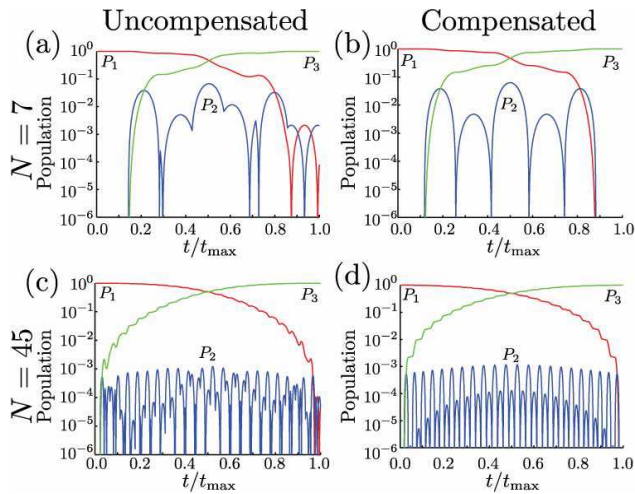}
\caption{Population through the linear protocol showing the effects of temporal compensation. (a) Uncompensated evolution with $N=7$, (b) compensated evolution with $N=7$, (c) uncompensated evolution for $N = 45$, and (d) compensated evolution for $N=45$.  In each case we have set $t_{\max} = \sum_{\xi} \tau(\xi)$, i.e. we have the set the total time to that of the compensated protocol.  Both (a,c) have residual populations in $\ket{1}$ and $\ket{2}$ at $t_{\max}$,  and also erratic transient population in $\ket{2}$ whereas (b,d) have complete population transfer with more regular transient $\ket{2}$ population.}
\label{fig:evolutioncorrected}
\end{figure}

\begin{figure}[tb!]
\includegraphics[width=0.7\columnwidth,clip]{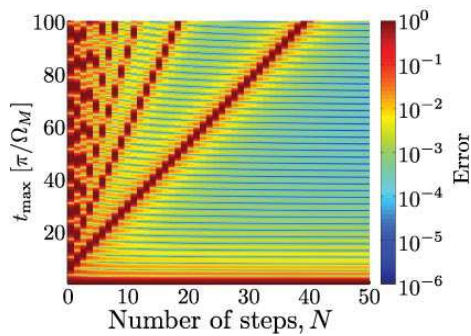}
\caption{Transport fidelity for the linear scheme with compensated pulse times.  This figure shows that the compensated pulses smoothens the function and gives rise to superior transport than the equally spaced protocol [c.f. Fig.~\ref{fig:p3-error}(b)].}
\label{fig:linnew}
\end{figure}

In Fig.~\ref{fig:evolutioncorrected} we compare the evolution through DAP for the linear scheme without correction to the pulse time (i.e. uniform pulse length) [for $N=7$ in Fig.~\ref{fig:evolutioncorrected}(a) and $N=45$ in Fig.~\ref{fig:evolutioncorrected}(c)] and with the correction applied [for $N=7$ in Fig.~\ref{fig:evolutioncorrected}(b) and $N=45$ in Fig.~\ref{fig:evolutioncorrected}(d)].  Populations are shown on a logarithmic scale to highlight the transient population in $\ket{2}$.  As expected, the evolution is far more regular for the temporally compensated cases than the uniform pulse time results.  The regularisation of the evolution is also seen in Fig.~\ref{fig:linnew}, which shows transport fidelity as a function of $N$ and $t_{\max}$, and should be compared with the plots in Fig.~\ref{fig:p3-error}.  Note that the resonances of identity evolution have sharpened so as to be comparable to those of the sin/cos protocol.

\section{Conclusions}
Digital adiabatic passage, when applied to three state systems, is shown to approximate conventional adiabatic passage in the limit of large $N$.  Although this was expected, it is perhaps surprising that the adiabatic limit should be well approximated even for relatively few digital steps, albeit with certain restrictions on the total time of the protocol.  In addition to adiabatic-like behaviour, we also find `resonances' exhibiting poor transport, superimposed on the adiabatic-like evolution.  These resonances of poor transport arise due to evolution over some or all of the digital steps resulting in the identity.

We compared two schemes for effecting digital adiabatic passage: sin/cos and linear variation; for uniform pulse lengths and compensated pulse lengths.  Without pulse length compensation, the sin/cos scheme gave superior transport results.  The overall transport was generally higher fidelity as the adiabaticity parameter of the sin/cos scheme was less than that for the linear scheme, however also the resonances of poor transport were more clearly defined due to the fact that the eigenergies were constant throughout the protocol.  Our results show that the constant eigenenergies of the sin/cos scheme provide improvements to the more commonly investigated adiabatic passage schemes. 

Optimisation of the pulse length for the linear scheme improved the fidelity of this approach to that of the sin/cos approach.  Whilst non-uniform pulse lengths are more complicated, in some systems, it may be possible to employ finer control over timing than the tunnel matrix element, and hence improve the fidelity of digital adiabatic passage.   We assumed that the tunnel matrix elements were selected from a set of uniformly spaced levels.  However, our method may be generalised to cases where the tunnel matrix elements vary within any set of discrete levels.

Overall, our results show that adiabatic passage is robust against digitisation of the control parameters, a result that is not obvious from simple adiabatic analysis.  The implications of these results is that the class of systems that are amenable to adiabatic passage techniques is larger than previously thought.

\section*{Acknowledgements}
The authors would like to thank Andrew Dzurak, Ben Hess, Jared Cole and Anthony Hope for useful conversations.  ADG also acknowledges the ARC for financial support (DP0880466 and DP130104381).


\begin{thebibliography}{99}

\bibitem{bib:GRB+1988} U. Gaubatz, P. Rudecki, M. Becker, S. Schiemann, M. K\"{u}lz and K. Bergmann, Chem. Phys. Lett. \textbf{149}, 463 (1988)

\bibitem{bib:KTS2007} P. Kr\'{a}l, I. Thanopulos, M. Shapiro, \rmp \textbf{79}, 53 (2007).


\bibitem{bib:KR1998} M. N. Kobrak and S. A. Rice,  J. Chem. Phys. \textbf{109}, 1 (1998).

\bibitem{bib:CK1995} S. Chelkowski and G. N. Gibson, \pra \textbf{52}, R3417 (1995).

\bibitem{bib:FGG+2001} E. Farhi, J. Goldstone, S. Gutmann, and M. Sipser, arXiv:quant-ph/0001106.

\bibitem{bib:ELR+2004} K. Eckert, M. Lewenstein, R. Corbal\'{a}n, G. Birkl, W. Ertmer, and J. Mompart \pra \textbf{70}, 023606 (2004).

\bibitem{bib:GCH+2004} A. D. Greentree, J. H. Cole, A. R. Hamilton, and L. C. L. Hollenberg \prb \textbf{70}, 235317 (2004).

\bibitem{bib:SB2004} J. Siewert and T. Brandes, Adv. Solid State Phys. \textbf{44}, 181 (2004) .

\bibitem{bib:GKW2006} E. M. Graefe, H. J. Korsch and D. Witthaut, \pra \textbf{73}, 013617 (2006).

\bibitem{bib:P2006} E. Paspalakis, Opt. Commun. \textbf{258}, 30 (2006).

\bibitem{bib:LDO+2007} S. Longhi, G. Della Valle, M. Ornigotti, and P. Laporta \prb \textbf{76}, 201101(R) (2007).

\bibitem{bib:RCP+2008} M. Rab, J. H. Cole, N. G. Parker, A. D. Greentree, L. C. L. Hollenberg, and A. M. Martin, \pra \textbf{77}, 061602(R) (2008).

\bibitem{bib:OEO+2008} T. Ohshima, A. Ekert, D. K. L. Oi, D. Kaslizowski, L. C. Kwek, arXiv:quant-ph/0702019.

\bibitem{bib:OSF+2013} S. Oh, Y.-P. Shim, J. Fei, M. Friesen, and X. Hu, \pra \textbf{87}, 022332 (2013).

\bibitem{bib:SMM+2007} E. A. Shapiro, V. Milner, C. Menzel-Jones, and M. Shapiro \prl \textbf{99}, 033002 (2007).

\bibitem{bib:SMS2009} E. A. Shapiro, V. Milner, and M. Shapiro, \pra \textbf{79}, 023422 (2009).

\bibitem{bib:TGV2011} B. T. Torosov, S. Guerin, N. V. Vitanov, \prl \textbf{106}, 233001 (2011).

\bibitem{bib:RV2012} A. A. Rangelov and N. V. Vitanov \pra \textbf{85}, 043407 (2012).

\bibitem{bib:BS2010} S. Boixo and R. D. Somma, \pra \textbf{81}, 032308 (2010).

\bibitem{bib:PA2012} G. Porat and A. Arie, J. Opt. Soc. Am. B  \textbf{29}, 2901-2909 (2012).

\bibitem{bib:CGH+2008} J. H. Cole, A. D. Greentree, L. C. L. Hollenberg, and S. Das Sarma, \prb \textbf{77}, 235418 (2008).

\bibitem{bib:RPC+2009} R. Rahman, S. H. Park, J. H. Cole, A. D. Greentree, R. P. Muller, G. Klimeck, and L. C. L. Hollenberg, \prb \textbf{80}, 035302 (2009).

\bibitem{bib:CH1990} C. E. Carroll and F. T. Hioe \pra \textbf{42}, 1522 (1990).

\bibitem{bib:LS1996} T. A. Laine and S. Stenholm, \pra \textbf{53}, 2501 (1996).

\bibitem{bib:Mes} A. Messiah, \textit{Quantum Mechanics} (North-Holland, Amsterdam,
1965), Vol. 2.

\bibitem{bib:DP1976} J. P. Davis and P. Pechukas, J. Chem. Phys. \textbf{64}, 3129 (1976).

\bibitem{bib:VCM+2012} N. Vogt, J. H. Cole, M. Marthaler, and G. Sch\"{o}n, \prb \textbf{85}, 174515 (2012).

\bibitem{bib:VL1998} L. Viola and S. Lloyd \pra \textbf{58}, 2733 (1998).

\end{thebibliography}
\end{document}